\documentclass[10pt]{article}
\usepackage[centertags]{amsmath}
\newcommand{\be}{\begin{equation}}
\newcommand{\ee}{\end{equation}}
\newcommand{\bea}{\begin{eqnarray}}
\newcommand{\eea}{\end{eqnarray}}
\usepackage[centertags]{amsmath}
\usepackage{amsfonts}
\begin{document}
\title{\bf Supersymmetry and R-symmetry breaking \\ in models with non-canonical K\"{a}hler potential}
\author{\normalsize L.~G.~Aldrovandi$^{a \, *}$ and D.~Marqu\'es$^a$\thanks{Associated with
CONICET}
\\
{\small\it $^a$Departamento de F\'{\i}sica-IFLP, Universidad Nacional de
La Plata, C.C. 67, 1900 La Plata, Argentina}}

\maketitle
\begin{abstract}
We analyze several aspects of R-symmetry and supersymmetry breaking
in generalized O'Raifeartaigh models with non-canonical K\"{a}hler
potential. Some conditions on the K\"{a}hler potential are derived in
order for the non-supersymmetric vacua to be degenerate. We
calculate the Coleman-Weinberg (CW) effective potential for
general quiral non-linear sigma models and then study
the 1-loop quantum corrections to the pseudo-moduli space. For
R-symmetric models, the quadratic dependence of the CW potential
with the ultraviolet cutoff scale disappears. We also show that the
conditions for R-symmetry breaking are independent of this scale and
remain unchanged with respect to those of canonical models. This is,
R-symmetry can be broken when generic R-charge assignments to the
fields are made, while it remains unbroken when only fields with
R-charge $0$ and $2$ are present. We further show that these models
can keep the runaway behavior of their canonical counterparts and also
new runaway directions can be induced. Due to the runaway directions, the non-supersymmetric
vacua is metastable.
\end{abstract}

\section{Introduction}

The first proposals for dynamical supersymmetry (SUSY) breaking
appeared to be rather non-generic (for a review see
\cite{Giudice:1998bp}), because several classic constrains
\cite{Witten:1982df} hardly restricted model building. These
constraints are removed if metastability for the vacua is accepted,
this giving rise to new possibilities for model building. In fact,
by demonstrating the metastable structure of the vacuum in massive
${\cal N} = 1$ SQCD (ISS model), it was shown in
\cite{Intriligator:2006dd} that metastable dynamical supersymmetry
breaking is much more generic and simpler than was previously
thought. In the low energy limit of this model (and other SUSY gauge
theories), O'Raifeartaigh-type models \cite{O'Raifeartaigh:1975pr}
arise naturally and dynamically, and are therefore appealing
candidates for the hidden sector of low-scale supersymmetric
theories \cite{Cheung:2007es}. Increasing efforts have been made to characterize common
aspects of supersymmetry breaking in these models. Among the many
common features that are sheared by these generic theories with
metastable vacua one should mention

${\bullet}$ Supersymmetry breaking and R-symmetry are connected. It
was shown in \cite{Nelson:1993nf} that the existence of an
R-symmetry is a necessary condition for supersymmetry breaking, and
a spontaneously broken R-symmetry is sufficient. When the theory is
not R-symmetric, it can contain supersymmetric vacua.

${\bullet}$ Runaway
directions are in general present implying that the SUSY-breaking
minima is only local \cite{Ferretti:2007ec}.
This vacua can be taken to be sufficiently long lived. When the
R-symmetry is softly broken supersymmetric vacua can appear, but
they can be pushed far away in field space
\cite{Intriligator:2007py}, \cite{Abe:2007ax}.

${\bullet}$ The supersymmetry breaking vacua is degenerate at
tree-level \cite{Ray:2006wk}. It corresponds to a pseudo-moduli space
further lifted by quantum corrections. If an exact R-symmetry is
present at the classical level, the corrections determine whether R-symmetry is broken or
not through the 1-loop effective Coleman-Weinberg (CW) potential
\cite{Coleman:1973jx}.

${\bullet}$ R-symmetry can be broken at the quantum level when the
R-charge assignment to the fields is generic \cite{Shih:2007av}.
When fields with only R = 0, 2, are present in the theory,
R-symmetry cannot be broken.

${\bullet}$  The effective potential includes a quartic divergent
term proportional to $\Lambda^4\ {\rm STr}\ 1$, and a quadratic
divergent term proportional to $\Lambda ^2\ {\rm STr}\ {\cal M}^2$,
with $\Lambda$ the UV cut-off scale. Both vanish in renormalizable
supersymmetric theories.

These items are only shared by theories with canonical K\"{a}hler
potentials, i.e. those in which the K\"{a}hler metric is the identity.
For instance, it is not necessarily true that models with
non-canonical K\"{a}hler posses non-supersymmetric degenerate vacua
manifolds. Moreover, the theories are not renormalizable and
quadratic and quartic divergent terms appear in the CW effective potential,
making them very sensitive to variations of the energy scale. These
are (some of) the reasons why generic aspects of supersymmetry
breaking in these kind of models have not been deeply studied.

This paper is devoted to study several aspects of
O'Raifeartaigh-type models with non-canonical K\"{a}hler potentials.
Interestingly, not so restrictive conditions have to be imposed on
the K\"{a}hler metric, in order for the theory to share the above
mentioned properties. We start showing some conditions on the K\"{a}hler
potential that imply degenerate vacua at tree level. We then analyze
the properties of this vacua, and show that its characterization is
completely analogous to that of the canonical K\"{a}hler models. We also
show sufficient conditions for these theories to have runaway
directions, making the non-supersymmetric minima metastable.

There are many situations in which non-canonical K\"{a}hler potentials
arise. The theories we consider are conceived as low-energy
effective theories of more fundamental renormalizable theories.
Loops of modes from these high energy theories induce effective
K\"{a}hler potentials \cite{Intriligator:2006dd}. It is worth noticing that, since supergravity
corrections are neglected, the scales associated to the higher order terms in the Kahler potential are assumed to be much smaller than the Planck scale. Non-minimal K\"{a}hler
potentials also arise in finite temperature and supergravity
theories and were studied in the context of metastability in
\cite{Abe:2007ax},\cite{Craig:2006kx}.

The outline of this paper is as follows. In section 2, we derive
sufficient conditions on the K\"{a}hler, in order for the scalar
potential to have a tree-level degenerate non-SUSY vacua. The most
general K\"{a}hler potential satisfying these conditions for a
generalized O'Raifeartaigh model is constructed. This derivation is
done for a one-dimensional pseudo-moduli space, and we also
generalize the result to the case of higher dimensional
pseudo-moduli space. In section 3, the model is analyzed in more
detail and its main characteristics are discussed. Section 4 is
devoted to the analysis of R-symmetry breaking, based on the 1-loop
quantum lifting of the flat directions of the pseudo-moduli space.
In section 5 we study the non-canonical version of an O'Raifeartaigh
model introduced by Shih \cite{Shih:2007av}, providing an explicit
realization of the main results of this paper. In section 6 we
present a summary and a discussion of our results. Finally, we add
an appendix with the computation of the 1-loop CW effective
potential for a general supersymmetric non-linear sigma model.

\section{Degeneracy for non-canonical K\"{a}hler}

The non-SUSY vacua of renormalizable Wess-Zumino models always
consists of a tree-level pseudo-moduli space lifted by quantum
corrections. If the theory is R-symmetric, the lifting determines if
R-symmetry is broken or not. In general, when studying these
theories, one usually relaxes the condition of genericity on the
superpotential, in order to obtain a deeper understanding of the
SUSY breaking properties of concrete models. These models, although
non generic, are quite general (see for example the models of
\cite{Shih:2007av}), consisting in families of models sharing
similar properties.

When we turn the attention to non-renormalizable theories with
non-canonical K\"ahler potential, there need not be a moduli
parameterizing the vacua: non-canonical corrections to the canonical
K\"ahler potential lift the moduli space at tree level. As this
lifting depends exclusively on the form of the K\"ahler potential, it
is much more difficult to analyze general aspects of SUSY breaking in
non-renormalizable models.

However, we can relax the genericity condition on the K\"ahler
potential (we call ``generic'' to those K\"ahler potentials containing
all the terms consistent with the symmetries of the theory), and try
to look for families of K\"ahler potentials sharing SUSY breaking and
R-symmetry breaking properties.

In this paper, as a first step in analyzing general aspects of SUSY
and $U(1)_R$ breaking in non-renormalizable models, we focus on
those families of non-canonical models that share the properties of
their canonical counterparts, which were deeply studied and are well
understood. Then, as a first step in our analysis, we must look for
conditions on the K\"ahler potential in order for the theory to have a
pseudo-moduli space.

One can think of a further step in the study of non-canonical
models, as that in which the conditions we derive are relaxed,
implying a tree level lifting of the moduli. We shall also make very
brief comments about this possibility in this section, but only
superficially, leaving this for further research.

\subsection{Sufficient conditions for degenerate vacua}\label{sec2.1}

This section is devoted to find a set of sufficient conditions on
a general K\"{a}hler potential $K$ and  the superpotential $W$ that imply
degeneracy of the (supersymmetry-breaking) vacua.

Let us first review what happens in the case of a theory with
canonical K\"{a}hler potential $K=\phi^a\delta_{a\bar a}\bar\phi^{\bar
a}$ (here $a=1,...,N_{\phi}$ label $N_{\phi}$ chiral fields $\phi^a$). In this case one can show \cite{Ray:2006wk} that if the
potential $V = \bar W_{\bar a}\delta^{\bar a a}W_a$ admits a local
non-supersymmetric vacuum, then a set of
vacua with the same tree-level energy forming a (continuous) submanifold of the
field space necessarily exists. More in detail, from the conditions for a field configuration $\phi_0,\bar\phi_0$
to be a non-supersymmetric vacuum:

$\bullet~ W_a|_{\phi_0} \neq 0$

$\bullet~ \bar W_{\bar a} \delta^{\bar a a}\partial_a W_b|_{\phi_0,\bar\phi_0} =0$

$\bullet~$ $\delta V \geq 0$  at the leading order in the variations $\delta \phi^a,\delta \bar\phi^{\bar a}$  for
any $\delta \phi^a, \delta \bar\phi^{\bar a}$

$\!\!\!\!\!\!\!\!\!\!\!$ one can prove that
\be
    \bar W_{\bar a_1}\delta^{\bar a_1 a_1}...\bar W_{\bar a_n}
\delta^{\bar a_n a_n}\partial_{a_1...a_n}W_b
|_{\phi_0,\bar\phi_0}=0\ , \quad\quad\quad\quad\quad       \forall
\;n\geq 1\ . \ee
Clearly, this result implies that
\be
     V(\phi_0^a + z \bar W_{\bar a}\delta^{\bar a a},\bar\phi_0^{\bar a} + \bar z \delta^{\bar a a}W_{a}) = V(\phi^a_0,\bar\phi_0^{\bar
     a}) \ ,
\ee
for any complex $z$, and then the potential is degenerate at tree-level.

The latter theorem only holds for a canonical K\"ahler potential. In
fact,  the vacuum need not to be degenerate for a generic K\"ahler
potential, as can be easily verified through the following simple
counter-example presented in \cite{Intriligator:2006dd}. Consider
a theory containing a single chiral superfield $X$, with linear
superpotential with coefficient $f$
\be
    W= f X\ ,
\ee
and an arbitrary K\"{a}hler potential $K(X,\bar X)$. The scalar
potential is
\be
   V = (\partial_X \partial_{\bar X}K)^{-1}|f|^2\ .
   \label{pot}
\ee
Let us suppose that the K\"{a}hler potential $K$ is smooth. For
smooth $K$, the potential (\ref{pot}) is non-vanishing, and thus
there is no supersymmetric vacuum. It is also clear that the vacuum
is not necessarily degenerate. Consider, for instance, the behavior
of the system near a particular point, say $X \approx 0$. Let
\be
   K = \bar X X - c (\bar X X)^2 + ...
\ee
with positive $c$. Then there is a locally stable non-supersymmetric vacuum at $X = 0$ and no degeneracy at all.

In spite of this, we will show below that under certain assumptions on the K\"ahler
potential, the presence of a degenerate vacuum can be guaranteed.

First
of all, let us note that since we only consider regular (non-smooth $K$ signals the need to include
additional degrees of freedom at the singularity) and positive-definite K\"ahler metrics, the conditions a
given vacuum must satisfy to break supersymmetry depend only on the
superpotential $W$ and not on the form of the K\"ahler potential. That
is, if the metric $K_{ a\bar a}$ is regular and positive-definite, the
potential $V=\bar W_{\bar a} K^{\bar a a }W_a$ in the vacuum will
vanish if and only if the vector $W_a$ is null in that
vacuum.\footnote{An obvious corollary of this is that the connection
between $R$ symmetry and supersymmetry breaking pointed out by
Nelson and Seiberg \cite{Nelson:1993nf} is valid for any regular K\"ahler
potential.  If the K\"ahler (and therefore the theory) is not
R-symmetric, the N-S argument still holds as long as the
superpotential has R-charge R($W$) $\neq 0$.}

Let us now prove the theorem which guarantees the existence of a tree-level moduli space. We require the following conditions to be satisfied:
\bea
&&\bullet\; W_a|_{\phi_0}\vphantom{\frac 1 2} \neq 0 \label{condia1}\\
&& \bullet\; \bar W_{\bar a} \partial_b (K^{\bar a a}W_a)|_{\phi_0,\bar\phi_0} =0 \label{condib1}\\
&& \bullet\; \delta V = \sum_{n,p=0}^\infty \frac{1}{n!p!} \sum_{m=0}^n \sum_{q=0}^p
    \left(  \begin{array}{c}
          n\\
          m
         \end{array} \right)
    \left(  \begin{array}{c}
          p\\
          q
         \end{array} \right)\delta \bar\phi^{\bar a_1}...\delta \bar\phi^{\bar a_n} \delta \phi^{a_1}...\delta \phi^{a_p}
    \nonumber\\
         &&~~~~~~\quad\quad\quad\quad\quad\partial_{\bar a_1...\bar a_{n-m}} \bar W_{\bar a}\, \partial_{\bar a_{n-m+1}...\bar a_n  a_1...  a_{p-q}} K^{\bar a a}\,\partial_{a_{p-q+1}...a_p} W_a |_{\phi_0,\bar\phi_0}\;\;\geq 0 \quad\quad\quad \forall\, \delta\phi^a,\delta\bar\phi^{\bar a}\label{condic1}\\
&& \bullet\; \frac{d}{d\lambda} W_a|_{\phi_0}=0\label{condi1b}\\
&&\bullet\; \frac{d^m}{d\bar\lambda^m}\frac{d^n}{d\lambda^n}K^{\bar a a}|_{\phi_0,\bar\phi_0}=0\ , \quad\quad\quad \forall \;m,n \geq 0 \;/\;
    m+n>0 \ , \label{condi2}
\eea
where $d/d\lambda$ and $d/d\bar\lambda$ are defined by

\be
   \frac{d}{d\lambda}= \bar W_{\bar a} K^{\bar a a}|_{\phi_0,\bar\phi_0}\partial_a\ , \quad\quad\quad\quad
   \frac{d}{d\bar\lambda}=   K^{\bar a a}W_{a}|_{\phi_0,\bar\phi_0}\partial_{\bar a}\ .
\ee
The conditions (\ref{condia1})-(\ref{condic1}) imply that the field configuration $\phi_0,\bar\phi_0$ is a non-supersymmetric vacuum
of the theory. Concerning  conditions (\ref{condi1b}),(\ref{condi2}), their meaning become clearer by noticing that $d/d\lambda$ and $d/d\bar\lambda$ are the derivatives along the curve $\phi^a(\lambda)$,
$\bar\phi^{\bar a}(\bar \lambda)$ given by
\be
   \phi^a(\lambda)=\bar W_{\bar a} K^{\bar a a}|_{\phi_0,\bar\phi_0}(\lambda - \lambda_0) + \phi^a_0\ , \quad\quad\quad\quad\bar \phi^{\bar a}(\bar\lambda)=K^{\bar a a}W_{a}|_{\phi_0,\bar\phi_0}(\bar\lambda - \bar\lambda_0) + \bar\phi^{\bar
   a}_0\ .
   \label{curve}
\ee
Therefore, equation (\ref{condi2}) implies that $K^{\bar a a}$ is constant along the
curve given by eq.(\ref{curve})
\bea
    K^{\bar a a}(\phi(\lambda),\bar\phi(\bar\lambda)) = K^{\bar a a}(\phi_0,\bar\phi_0)\label{condika}\ ,
\eea
with $\lambda$ any complex number.\footnote{In other words, if we denote $\overrightarrow{U}$ to the tangent vector to the curve (\ref{curve}) (i.e. the vector field  with components $U^a = \bar W_{\bar a} K^{\bar a a}|_{\phi_0,\bar\phi_0}$), eq.(\ref{condi2}) implies that $\overrightarrow{U}$ is a Killing vector of the K\"{a}hler metric $\mathbb{K}$ when we restrict ourself to the curve $\phi(\lambda),\bar\phi(\bar\lambda)$, that is, the Lie derivative vanishes on this curve,
\be
   \pounds_{\overrightarrow{U}} { \mathbb{K}}
   |_{\phi(\lambda),\bar\phi(\bar\lambda)}=0\ .
\ee}

 Let us prove by recurrence that under the latter assumptions (\ref{condia1})-(\ref{condi2}) the potential
$V$ is always degenerate at tree-level. To do this we suppose, as a
recurrence condition, that for some non-zero integer $n$ we have
\be
    \frac{d^k }{d\lambda^k}W_a|_{\phi_0}= 0\ ,
    \quad\quad\quad\quad \forall\; 1\leq k \leq n\ .
    \label{condi3}
\ee
Let us then consider a variation of the fields $\phi^a$ around the
vacuum $\delta \phi^a = \bar W_{\bar a} K^{\bar a
a}|_{\phi_0,\bar\phi_0} \delta \lambda + \varphi^a
\delta\lambda^{n+1}$. The leading term of the variation of $V$ for
small $\delta \lambda$ must be positive whatever the choice of the
direction $\varphi^a$ is.

For $1 \leq k \leq n$, the $k$-th order of variation of $V$ in
$\delta\lambda$ reads
\be
   \delta^k V = \sum_{i=0}^k \frac{\delta\lambda^i\delta\bar\lambda^{k-i}}{i!(k-i)!}\frac{d^{k-i} }{d\bar\lambda^{k-i}}\bar W_{\bar a}
   K^{\bar a a}\frac{d^i }{d\lambda^i}W_a|_{\phi_0,\bar\phi_0}=0\ ,
\ee
by use of condition (\ref{condi2}) and recurrence relation
(\ref{condi3}). Furthermore, for $0 \leq k \leq n$, the $(n+k+1)$-th
order reads
\bea
   \delta^{n+k+1} V &=& \sum_{i=0}^{n+k+1} \frac{\delta\lambda^i\delta\bar\lambda^{n+k-i+1}}{i!(n+k-i+1)!}\frac{d^{n+k-i+1} }{d\bar\lambda^{n+k-i+1}}\bar W_{\bar a}
   K^{\bar a a} \frac{d^i }{d\lambda^i}W_a|_{\phi_0,\bar\phi_0}\nonumber\\
    &+& 2 {\rm Re} \left\{\sum_{i=0}^k \frac{\delta\lambda^{n+i+1}\delta\bar\lambda^{k-i}}{i!(k-i)!} \frac{d^{k-i} }{d\bar\lambda^{k-i}} \bar W_{\bar a} \varphi^b\partial_b (K^{\bar a a}\frac{d^i }{d\lambda^i}W_a)|_{\phi_0,\bar\phi_0}\right\}\nonumber\\
    &=& 2 {\rm Re}\left\{ \delta\lambda^{n+k+1}\left[ \frac{1}{(n+k+1)!} \bar W_{\bar a}K^{\bar a a} \frac{d^{n+k+1} }{d\lambda^{n+k+1}}W_a |_{\phi_0,\bar\phi_0}  \right.\right.\nonumber\\
    &+&\left.\left. \frac{1}{k!} \bar W_{\bar a}K^{\bar a a} \varphi^b\partial_b \frac{d^{k} }{d\lambda^{k}}W_a|_{\phi_0,\bar\phi_0}+\delta^{k 0}\bar W_{\bar a}\varphi^b\partial_b K^{\bar a a} W_a |_{\phi_0,\bar\phi_0}\right]
    \right\}=0\ .
\eea
The last term vanishes as a consequence of eqs.(\ref{condib1}),(\ref{condi1b}). The remaining terms must be all zero since, if one
of them were not, the leading order in $\delta\lambda$ would be of
the form ${\rm Re}(\delta\lambda^{n+k+1})$, which takes negative
values for some $\delta\lambda$. Hence,
\be
   \frac{1}{(n+k+1)!} \bar W_{\bar a}K^{\bar a a} \frac{d^{n+k+1} }{d\lambda^{n+k+1}}W_a|_{\phi_0,\bar\phi_0} +
\frac{1}{k!} \bar W_{\bar a}K^{\bar a a} \varphi^b\partial_b
\frac{d^{k} }{d\lambda^{k}}W_a|_{\phi_0,\bar\phi_0}=0\ ,
\quad\quad\quad\quad\forall\; 0 \leq k \leq n\ . \ee
Therefore, since $\varphi^a$ is an arbitrary vector, $\bar W_{\bar
a}K^{\bar a a} \partial_b \frac{d^{k}
}{d\lambda^{k}}W_a|_{\phi_0,\bar\phi_0}$ must be itself equal to
zero. Then, using that $\partial_b \frac{d^{k} }{d\lambda^{k}}W_a =
\partial_a \frac{d^{k} }{d\lambda^{k}}W_b$ and taking $k=n$ gives
the result
\be
    \frac{d^{n+1} }{d\lambda^{n+1}}W_a|_{\phi_0}= 0\ ,
\ee
so the recurrence condition is verified one step further. This, together with the fact that the recurrence
condition (\ref{condi3}) is true for $n = 1$, implies that
\be
    \frac{d^n }{d\lambda^n}W_a|_{\phi_0}= 0\ ,
    \quad\quad\quad\quad \forall\; n>0\ ,
\ee
and then $W_a$ is constant along the
curve eq.(\ref{curve}),
\bea
    W_a(\phi(\lambda),\bar\phi(\bar\lambda)) &=&
    W_a(\phi_0,\bar\phi_0)\ ,
\eea
with $\lambda$ any complex number. Since $K^{\bar a a}$ is also constant along this
curve (see eq.(\ref{condika})), it is trivial to check that the same happens with the potential $V=\bar W_{\bar a} K^{\bar a a}W_a$.

In summary, we have shown that when the conditions (\ref{condia1})-(\ref{condi2}) are satisfied, the potential $V$ is degenerate along a one-dimensional sub-manifold,
\be
    V(\phi(\lambda),\bar\phi(\bar\lambda)) = V(\phi_0,\bar\phi_0) \
    ,
\ee
with the curve $\phi(\lambda),\bar\phi(\bar\lambda)$ given by
\be\phi^a(\lambda)=\bar W_{\bar a} K^{\bar a a}|_{\phi_0,\bar\phi_0}(\lambda - \lambda_0) + \phi^a_0\ , \quad \quad\bar \phi^{\bar a}(\bar\lambda)=K^{\bar a a}W_{a}|_{\phi_0,\bar\phi_0}(\bar\lambda - \bar\lambda_0) + \bar\phi^{\bar
a}_0\ .\ee

\subsection{Models with one pseudomoduli}\label{sec2.2}

In this section we will use the results obtained in section \ref{sec2.1} to find the most general non-canonical K\"{a}hler consistent with a particular type of superpotentials recently analized by Shih \cite{Shih:2007av} in the case of canonical K\"{a}hler, which are a generalization of the original O'Raifeartaigh models. These R-symmetric superpotentials can be written as
\be
    W = f X + \frac{1}{2} (B_{ij} + X A_{ij}
    )\phi^i\phi^j , \label{Superpot}
\ee
where $f$ is a complex constant, and $A$ and $B$ are symmetric
complex matrices satisfying $det(B) \neq 0$ (see below). In order for the
superpotential to be R-symmetric, we require
\be
   A_{ij} \neq 0 \Rightarrow R(\phi^i) + R(\phi^j) = 0\ , \ \ \ \ \
   \ \ B_{ij} \neq 0 \Rightarrow R(\phi^i) + R(\phi^j) = 2\ .
   \label{consR}
\ee

In the case of canonical K\"{a}hler \cite{Shih:2007av}, supersymmetry is
broken in this model and a non-supersymmetric minimum $V_0 = |f|^2$
is given by
\be
   \phi^i = 0\ , \ \ \ \ \ \ X\ {\rm arbitrary}\ .
   \label{curve2}
\ee
Therefore, the field $X$ become a modulus parameterizing the vacua manifold. This planar direction is lifted by quantum corrections so $X$ is called a pseudomodulus.

Let us see now how the K\"{a}hler potential should be in order to
maintain the degeneracy at tree level along the curve
(\ref{curve2}). From condition (\ref{condi2}) (or, equivalently,
condition (\ref{condika})), the components $K^{\bar a a}$ must be
constant along the curve, and then
\be
    \partial_X K_{ a\bar a}(X,\phi^i=0)=0\ ,\quad\quad\quad\quad \partial_a K_{X\bar a
    }(X,\phi^i=0)=0\ .
    \label{con1}
\ee
Concerning condition (\ref{condi1b}), this can be written as
\be
    K^{\bar X i}(B_{ij} + X A_{ij})=0\ .
\ee
As shown by Shih \cite{Shih:2007av}, the constraints due to R-symmetry (\ref{consR}) imply that $det(B+XA)=det(B)\neq 0$. Therefore,
$K^{\bar X i}(X,\phi^i=0)$ must vanish, leading to
\be
    K_{X \bar i}(X,\phi^i=0) =0\ ,\quad\quad\quad\quad K_{i \bar X}(X,\phi^i=0)
    =0\ .
    \label{con2}
\ee
The last condition to impose, coming from eq.(\ref{condib1}), is $W_{\bar a}\partial_b K^{\bar a a} W_a (X,\phi^i=0) = |f|^2 \partial_b K^{\bar X X} (X,\phi^i=0)=0$. However, it is easy to show that in our case this condition is already implied by (\ref{con1}) and (\ref{con2}). From $K^{\bar X a}K_{a \bar X} =1$ we obtain
\be
   \partial_b K^{\bar X X}K_{X \bar X}+\partial_b K^{\bar X i}K_{i \bar X}+ K^{\bar X X}\partial_b K_{X \bar X}+ K^{\bar X i}\partial_b K_{i \bar
   X}=0\ .
\ee
When we evaluate this equation in $\phi^i=0$, the second and fourth
terms vanish due to (\ref{con2}), while the third one due to
(\ref{con1}), this  leaving us with the desired result.

In summary, the conditions we have to impose on the K\"{a}hler potential
to have degeneracy in $\phi^i=0$ for arbitrary $X$ are
\be
    \partial_{\bar a}\partial_{aX} K(X,\phi^i=0)=0\ ,\quad\quad\quad\quad \partial_{\bar i}\partial_{X}
    K(X,\phi^i=0)=0\ .
    \label{connn}
\ee
If we consider an expansion of $K$ in powers of $X$ and $\bar X$
\be
   K(X,\phi^i,\bar X,\bar\phi^{\bar j}) = \sum_{m,n =0}^\infty f_{mn}(\phi^i,\bar\phi^{\bar j})X^m\bar X^n,\quad\quad\quad\quad
   f_{nm}=(f_{mn})^*\ ,
\ee
then, conditions (\ref{connn}) can be expressed as
\bea
    &&\partial_{\bar j} f_{m0}(0) = \partial_i\partial_{\bar j} f_{m0}(0) = 0 \quad\quad\quad\quad \quad\quad\quad\quad\quad\quad\quad\quad \quad\quad\ \  m>0\nonumber\\
    && f_{11}(0) = {\rm const.}\ , \quad\quad \partial_i f_{11}(0) =\partial_{\bar j} f_{11}(0) = \partial_i\partial_{\bar j} f_{11}(0) = 0\nonumber\\
    && f_{mn}(0) = \partial_i f_{mn}(0) =\partial_{\bar j} f_{mn}(0) = \partial_i\partial_{\bar j} f_{mn}(0) = 0
    \quad\quad\quad\quad\ \ m,n\geq 1, m+n>2\ .
    \label{partial}
\eea The simplest example of K\"{a}hler potential leading to the
required degeneracy consists in imposing that equation
(\ref{partial}) be valid not only in $\phi^i=0$ but for any
$\phi^i$. In this case, $K$ can be written as
\be
   K = c X \bar X + C(\phi^i, \bar \phi^{\bar j})\
   ,\label{resu}
\ee
with $c$ any real constant (that can trivially be taken to $1$ by a
rescaling of $X$).

\subsection{Models with more pseudo-moduli}

Based on the result eq.(\ref{resu}), it is easy to propose a model possessing several pseudomoduli. An obvious generalization
of the non-canonical K\"{a}hler potential (\ref{resu}) is given by
\be K = X^\alpha \delta_{\alpha \bar \alpha} \bar X^{\bar\alpha} +
C(\phi^i, \bar \phi^{\bar j})\ ,\label{Kahler}\ee
where $\alpha, \beta, ... = 1,...,N_X$ label the $N_X$ fields
$X^\alpha$ that appear in $K$ in a canonical form, while $i, j, ...
= 1, ..., N_\phi$ label the $N_\phi$ fields $\phi^i$ with
non-canonical structure. Thus, the K\"{a}hler potential (\ref{Kahler})
defines a (regular and positive defined) metric in field space with
matricial form
\be K_{a\bar a} = \partial_a\partial_{\bar a} K = \left(\begin{matrix} \delta_{\alpha \bar \alpha} & 0\\ 0 &
C_{i \bar i}
\end{matrix}\right)\ , \label{KMatriz}\ee
where we have defined the $N_\phi \times N_\phi$ matrix
\be C_{i \bar j} = \partial_i\partial_{\bar j} C \ .\label{MatC}\ee

Concerning the superpotential, inspired in the generalized
O'Raifeartaigh models, we consider superpotentials with the form
\cite{Ferretti:2007ec}
\be W  = f^\alpha X^\alpha + \frac{1}{2} (B_{ij} + X^\alpha A^\alpha_{ij})\phi^i\phi^j \
,\label{SuperpotGeneral}\ee
for which the scalar potential $V =\bar W_{\bar a} K^{\bar a a} W_a$, reads
\be V =\ |f^\alpha + \frac{1}{2} A^\alpha_{ij}\phi^i\phi^j|^2 + \bar \phi^{\bar i}(\bar B + \bar X^{\bar \alpha} \bar A^{\bar \alpha})_{\bar i\bar j}C^{\bar j j}(B + X^\alpha A^{\alpha})_{ji}\phi^i\ .\label{GenericPot}\ee

In order for these models to be R-symmetric, $K$ must be R-symmetric
and the R-charges of the fields should be such that
\be
   R(X^\alpha)=2\ , \ \ \ \ \
   \ \ A^\alpha_{ij} \neq 0 \Rightarrow R(\phi^i) + R(\phi^j) = 0\ , \ \ \ \ \
   \ \ B_{ij} \neq 0 \Rightarrow R(\phi^i) + R(\phi^j) = 2\ .
   \label{consR2}
\ee

Analogously to the case of the superpotential (\ref{SuperpotAB}),
these models break supersymmetry. In fact, the equations for a
supersymmetry vacuum are
\bea
    f^\alpha + \frac{1}{2} A^\alpha_{ij}\phi^i\phi^j &=&0\ \nonumber\\
    (B_{ij} + X^\alpha A^\alpha_{ij})\phi^j &=&0 \ .
    \label{ecu}
\eea
Similarly to the case of one pseudomoduli, conditions (\ref{consR2}) imply that $det(B+X^\alpha A^\alpha)=det(B)\neq 0$. Therefore, equations (\ref{ecu}) are not compatible. Besides, it is clear that a non supersymmetric minimum $V = |f_\alpha|^2$
appears at
\be
   \phi^i = 0\ , \ \ \ \ \ \ X^\alpha \ {\rm arbitrary}\ ,
\ee
implying that the fields $X^\alpha$ are
pseudomoduli parameterizing the vacua manifold.

In this general case, we have a $N_X$-dimensional manifold of
non-supersymmetric vacua, parameterized by $X^\alpha$. Then, in
order to generalize this K\"ahler potential, we can think of a more
general dependence on the $X^\alpha$ fields
\be K = k(X^\alpha, \bar X^{\bar\alpha}) + C(\phi^i, \bar \phi^{\bar
j})\ ,\ee
with the corresponding tree-level lifting
\be V_{\phi = 0}(X^\alpha, \bar X^{\bar\alpha}) = f_\alpha k^{\alpha
\bar \alpha} f_{\bar \alpha}\ ,\ee
when $k^{\alpha \bar \alpha} \neq \delta^{\alpha\bar \alpha}$. This
is nothing but the generalization of (\ref{pot}), which is the
1-dimensional case. This ``tree-level moduli lifting'' case might be
studied on general grounds by a classification of the different
$k^{\alpha \bar \alpha}$ metrics (some examples in the 1-dimensional
case were computed in \cite{Intriligator:2007cp}). Moreover, this
would be a starting point to study new terms mixing the $X$ and
$\phi$ fields. In this paper we do not consider these
possibilities.

\section{O'Raifeartaigh models}

In this section we consider quiral models recently introduced in
\cite{Shih:2007av}, which are a generalization of the original
O'Raifeartaigh model \cite{O'Raifeartaigh:1975pr}. We have already
considered them in Section \ref{sec2.2}, but we define them again in
order for this section to be self-contained. The superpotential for
this theory is
\be W = f X + \frac{1}{2} B_{ij}\phi^i\phi^j + \frac{1}{2}X A_{ij}
\phi^i\phi^j\ , \label{SuperpotAB}\ee
where $f$ is a complex constant and $A$ and $B$ are symmetric
complex matrices. The matrix $B$ satisfies $det(B) \neq 0$ and $A$
and $B$ have non-zero entries only when
\be A_{ij} \neq 0 \Rightarrow R(\phi^i) + R(\phi^j) = 0\ , \ \ \ \ \
\ \ B_{ij} \neq 0 \Rightarrow R(\phi^i) + R(\phi^j) = 2\ ,\ee
so $W$ has a definite R-charge $R(W) = 2$. The susy vacua conditions
for this theory read
\begin{eqnarray}
f + \frac{1}{2} A_{ij} \phi^i \phi^j &=& 0\label{SusyCond1}\\
(X A + B)_{ij} \phi^j &=& 0\ .\label{SusyCond2}
\end{eqnarray}
Because of R-symmetry, $A$ and $B$ adopt the following matricial
form
\be A = \left(\begin{matrix} 0 & & & & A_1 &  & 0
\\ & & \ldots & & & & &
\\ A_1^T & & & & \\  & & & & & \\ 0 & & & & & & 0 \end{matrix}\right)\ , \ \ \ \ \ \  B = \left(\begin{matrix} 0& & & & B_1 \\ & & & B_2 &
\\ & & \ldots & & \\ & B_2^T & & & \\ B_1^T & & & & 0\end{matrix}\right) \label{ABMatrices}\ee
in some field basis. As shown by Shih \cite{Shih:2007av}, this
particular structure for the matrices implies $det(X A + B) =
det(B)$. Then, as we have taken this to be non-zero, the only
solution for (\ref{SusyCond2}) is $\phi_i = 0 \ \forall i$, so
(\ref{SusyCond1}) can never be satisfied. Susy is therefore broken
in this model, and a non-supersymmetric minimum is given by
\be \phi^i = 0\ , \ \ \ \ \ \ X\ {\rm arbitrary}\ , \ \ \ \ \ \ V_0
= |f|^2 \ . \ee

The above considerations are independent of the K\"{a}hler potential.
For $K$ we consider that obtained in (\ref{resu}), which can be
written without loss of generality as
\bea K &=& \bar X X + C(\phi, \bar \phi) \nonumber \\
C(\phi, \bar \phi) &=& \phi^i C_{i\bar j} \bar\phi^{\bar j} + \dots
\ .\label{KahlerPosta}\eea
Here we have taken the $c$ parameter in (\ref{resu}) to be $c = 1$
by rescaling the field $X$. $C^{i \bar j}$ is an hermitic matrix
that satisfies $C^{i \bar j} \neq 0 \Rightarrow R(\phi^i) + R(\bar
\phi^{\bar j}) = 0$, and ``$\dots$'' are cubic or higher terms. In
the basis in which $A$ and $B$ take the form (\ref{ABMatrices}), $C$
has a diagonal-block form with blocks of fields having the same
R-charge.

It is easy to see that performing a change of the field basis (not
necessarily a unitary transformation), the quadratic part of the
K\"{a}hler can be taken to have a canonical form, leaving the
superpotential with the same structure as in (\ref{SuperpotAB})
\bea K &=& \bar X X + D(\phi, \bar\phi)\ , \label{KahlerC}\\
D(\phi, \bar\phi) &=& \phi^i \delta_{i\bar j} \bar\phi^{\bar j} +
\dots\ ,\\
W &=& f X + \frac{1}{2} M_{ij}\phi^i\phi^j + \frac{1}{2}X N_{ij}
\phi^i\phi^j\ . \label{SuperpotNM} \eea
Here we have written the transformed fields and introduced new
matrices $N$ and $M$, which are generic symmetric matrices with the
same form of $A$ and $B$ in (\ref{ABMatrices}) respectively. Then,
after this change of basis, we are left (in a neighborhood of the
$\phi^i = 0$ vacua) with a theory with superpotential
(\ref{SuperpotNM}) and canonical K\"{a}hler potential.

Although we have diagonalized the quadratic dependence of the K\"ahler potential, one can not get rid of its curvature and then, those properties depending on cubic and higher order terms will change. Interestingly, the stability of the $\phi^i = 0$ pseudo-moduli space in the $\phi^i$ direction is not affected by cubic or higher order terms. The reason for this is that the mass squared matrix for bosons (see eq.(\ref{MB2}) in appendix) in this vacua only depends on $K^{a\bar a}$ and not on its derivatives.

One important feature that arises when the K\"{a}hler is non-minimal is
that the mass squared matrices (see eqs.(\ref{MB2}),(\ref{MF2}) in
the appendix) get modified in such a way that their eigenvalues
split, even at tree level. The so-called supertrace theorem \cite{Ferrara:1979wa} generically implies the existence of a supersymmetric particle lighter than its ordinary partner, and then the paradigm for constructing realistic SUSY theories is to assume that the SUSY-breaking sector has no renormalizable tree level couplings with the observable sector. The latter theorem follows from the properties of renormalizability that force the kinetic terms to have a canonical form. This is not our case, as we are considering effective low-energy theories which not necessarily have a canonical K\"ahler potential, this leading  to the important phenomenological consequence mentioned above of mass splitting at tree-level. The mass-squared matrices of this model
(\ref{SuperpotAB}),(\ref{KahlerPosta}) in the vacuum $\phi^i = 0$
are
\begin{eqnarray}
{\cal M}_F^2 &=& (\hat B + X \hat A)^2 \nonumber \\
{\cal M}_B^2 &=& (\hat B + X \hat A)\ \hat C\ (\hat B + X \hat A) +
f \hat A\ ,\label{MByMF}
\end{eqnarray}
where we have defined
\be \hat A = \left(\begin{matrix}0 & A \\ A^\dag & 0
\end{matrix}\right)\ , \ \ \ \ \ \hat B = \left(\begin{matrix}0 & B \\ B^\dag & 0
\end{matrix}\right)\ , \ \ \ \ \ \hat C = \left(\begin{matrix} C^{-1} & 0 \\ 0 & C^{-1}
\end{matrix}\right)\ , \label{ABTilde}\ee
being $A$ and $B$ the matrices (\ref{ABMatrices}), and $C$ the
matrix defined in (\ref{MatC}) which has a diagonal block form with
blocks of fields having the same R-charge.

Following Ferretti's approach \cite{Ferretti:2007ec}, we now look
for possible restrictions to the K\"{a}hler potential by demanding this
theory to posses runaway directions. As (\ref{SusyCond1}) and
(\ref{SusyCond2}) are incompatible, we look for a compatible subset
of equations. In fact, classifying (\ref{SusyCond2}) acording to
their R-charge we have
\begin{eqnarray}
(X A + B)_{ij} \phi^j &=& 0\ , \ \ \ \ \ \ R(\phi^i) < 2 \label{SusyCond2a}\\
(X A + B)_{kj} \phi^j &=& 0\ , \ \ \ \ \ \ R(\phi^k) = 2 \label{SusyCond2b}\\
(X A + B)_{mj} \phi^j &=& 0\ , \ \ \ \ \ \ R(\phi^m) > 2\ ,
\label{SusyCond2c}
\end{eqnarray}
and it was shown in \cite{Ferretti:2007ec} that for generic R-charge
assignments it is always possible to solve
(\ref{SusyCond1})-(\ref{SusyCond2a})-(\ref{SusyCond2b}). We call the
solutions to these equations $X', \phi'^i$. The potential for this
particular configuration of fields reads
\be V_0' = \sum_{R(\phi^m) > 2\ \& \ R(\phi^n) > 2} C'^{\bar m
n}\left(X' A + B)_{ni}\right(\bar X'\bar A +\bar B)_{\bar j\bar m}
\phi'^i \bar \phi'^{\bar j}\ . \ee
Interestingly, by looking at
(\ref{SusyCond1})-(\ref{SusyCond2a})-(\ref{SusyCond2b}), we see that
a continuously connected range of solutions parameterized by a
parameter $\delta$ is obtained from every solution $X', \phi'^i$
\be \phi^i(\delta) = \delta^{-R(\phi^i)} \phi'^i \ , \ \ \ \ \ \
X(\delta) = \delta^{-2} X'\ . \ee
For these fields, the potential now reads
\begin{eqnarray}
V_0(\delta) &=& \sum_{R(\phi^m) > 2\ \& \ R(\phi^n) > 2}
(C(\delta))^{\bar m n}(X(\delta) A + B)_{ni}(\bar X(\delta)\bar A
+\bar B)_{\bar j\bar m}
\phi^i(\delta) \bar \phi^{\bar j}(\delta)\\
&=& \sum_{R(\phi^m) > 2\ \& \ R(\phi^n) > 2} (C(\delta))^{\bar m
n}(X' A + B)_{ni}(\bar X'\bar A +\bar B)_{\bar j\bar m} \phi'^i \bar
\phi'^{\bar j} \delta^{-4 + R(\phi^m) + R(\phi^n)}\ .\nonumber
\end{eqnarray}
This potential slopes to zero when
\be \lim_{\delta \to 0}\delta^{R(\phi^m) + R(\phi^n)-4}
(C(\delta))^{\bar m n} =0 \ , \ \ \ \ \ \ \forall n, m / R(\phi^m)
> 2\ \& \ R(\phi^n) > 2 \ . \label{RunawayCond}\ee
This doesn't seem a hard restriction provided that already
$\delta^{R(\phi^m) + R(\phi^n)-4} \to 0$ when $\delta \to 0$. We
have found a sufficient condition on the K\"{a}hler potential in order
for the theory to have runaway behavior. Moreover, the K\"{a}hler can
induce runaway behavior, even if the canonical theory has no runaway
directions.

Recently, strongly convincing arguments have been given that we happen to live in a metastable vaccum \cite{Intriligator:2006dd}. Thus, in order to construct
viable phenomenological models, besides the SUSY-breaking minimum, these models must have runaway directions and/or supersymmetric vacua. Moreover, the notion of meta-stable states is meaningful only when they are parametrically long lived since, phenomenologically, we would like the lifetime of our meta-stable state to be longer than the age of the Universe. It is therefore important for us to have the possibility of modifying the landscape of vacua by adjusting the parameters of the K\"ahler potential, since in this way one can guarantee the longevity of the meta-stable state.

\section{R-symmetry breaking}

For the O'Raifeartaigh models of the previous sections
(\ref{SuperpotAB}),(\ref{KahlerPosta})
\be W = f X + \frac{1}{2} B_{ij}\phi^i\phi^j + \frac{1}{2}X A_{ij}
\phi^i\phi^j\ ,\ \ \ \ \ \ K = \bar X X + C(\phi, \bar\phi)\ , \ee
where X is the coordinate parameterizing the pseudomoduli space.
The fact that $X$ is a coordinate of the one-dimensional vacua
manifold requires analysis beyond tree-level. Thus, we expect that
radiative corrections to the scalar potential will determine the
vacuum properties dynamically. Moreover, these corrections must
respect the symmetries of the original theory, so we can already
anticipate their form
\be V_{\rm eff}(|X|^2) = V_0 + m_{X}^2 |X|^2 + {\cal O}(|X|^4)\ .\ee
It has an extremum at $X = 0$ so it lifts the classical vacuum
degeneracy. Moreover, as it is shown below, $m_{X}^2$ can take values which are not necessarily positive.

The first order in the loop expansion of $V_{\rm eff}$
is given by the formula (see eq.(\ref{EffectivePot}) in Appendix)
\be V^{(1)}_{\rm eff} = \frac{1}{64 \pi^2} {\rm Tr}\left(
\tilde{\cal M}_B^4 \left[\log\left(\frac{\tilde {\cal
M}_B^2}{\Lambda^2}\right) - \frac{1}{2}\right] - \tilde {\cal M}_F^4
\left[\log\left(\frac{\tilde {\cal M}_F^2}{\Lambda^2} \right) -
\frac{1}{2} \right] + \frac{\Lambda^2}{2} \left(\tilde {\cal M}_B^2
- \tilde {\cal M}_F^2\right)\right) \label{OneLoopEff}\ ,\ee
where
\be \tilde {\cal M}_F^2 = K^{-1/2} {\cal M}_F K^{-1} {\cal M}_F
K^{-1/2}\ , \ \ \ \ \ \ \tilde {\cal M}_B^2 = K^{-1/2} {\cal M}_B^2
K^{-1/2} \ . \ee
There is also an additional term proportional to $\Lambda^4$, which we omit here because it is constant. Our aim
is to derive a general formula for $m_X^2$ in the one-loop
approximation as was done in \cite{Shih:2007av} but in the
non-canonical model proposed in the previous sections. This will
tell us wether the $X$ filed acquires a VEV or not. If it does
($m_X^2 < 0$) then, as $R(X) = 2$, R-symmetry is broken. Otherwise,
R-symmetry remains unbroken in this vacuum.

For the O'Raifeartaigh models
the tilde - mass matrices read
\begin{eqnarray}
\tilde {\cal M}_F^2 &=& {\hat C}^{1/2}\ (\hat B + X \hat A)\ \hat C\ (\hat B + X \hat A)\ {\hat C}^{1/2} \nonumber \\
\tilde {\cal M}_B^2 &=& {\hat C}^{1/2}\ (\hat B + X \hat A)\ \hat C\
(\hat B + X \hat A)\ {\hat C}^{1/2} + f \hat A\ ,\label{MByMFtilde}
\end{eqnarray}
where $\hat A$, $\hat B$ and $\hat C$ have been defined in
(\ref{ABTilde}), and the following identities hold
\be {\rm Tr} (\tilde {\cal M}_B^2 - \tilde {\cal M}_F^2) = 0\ , \ \
\ \ {\rm Tr} \frac{\partial^2}{\partial X^2} (\tilde {\cal M}_B^4 -
\tilde {\cal M}_F^4)\mid_{X = 0} = 0\ .\label{nonquadratic} \ee
This is a very interesting result because it implies that one-loop
corrections to the scalar potential are not quadratic, but
logarithmic in $\Lambda$. And also, $m_X^2$ is independent of
$\Lambda$. These two features are always true (independently of the
superpotential) in canonical K\"{a}hler models. Here, it is true due to
the form these matrices adopt because of the R-symmetry.

Then, in this case, the effective potential can be written as
\be V_{\rm eff}^{(1)} = -\frac{1}{32\pi^2}{\rm Tr} \int^\Lambda_0 dv
\ v^5\ \left(\frac{1}{v^2 + \tilde {\cal M}_B^2} - \frac{1}{v^2 +
\tilde {\cal M}_F^2}\right)\ ,\label{VeffIntegral}\ee
and we can substitute (\ref{MByMFtilde}) in (\ref{VeffIntegral}) to
obtain an expression for $m_X^2 = \frac{1}{2} \frac{\partial^2
V^{(1)}_{\rm eff}}{\partial X^2} \mid_{X = 0}$, which is
$\Lambda$-independent
\begin{eqnarray}
m_X^2 = \frac{1}{16 \pi^2}{\rm Tr}\int^\infty_0 dv\ v^3\
&&\!\!\!\!\!\!\!\!\!\!\!\!\!\left[\frac{1}{v^2 + {\cal B}^2 + f\hat
A}\left({\cal A}^2  - \frac{1}{2}\left\{ {\cal A}, {\cal B} \right\}
\frac{1}{v^2 + {\cal B}^2 + f\hat A}\left\{
{\cal A},  {\cal B} \right\}\right) \right.\nonumber\\
&&\!\!\!\!\!\!\!\!\!\!\!\!-\left. \frac{1}{v^2 + {\cal
B}^2}\left({\cal A}^2  - \frac{1}{2}\{{\cal A}, {\cal B}\}
\frac{1}{v^2 + {\cal B}^2}\{{\cal A}, {\cal B} \}\right)\right]\
.\label{ExpreVeff}
\end{eqnarray}
Here we have integrated by parts and defined
\be {\cal A} = \hat C^{1/2} \hat A \hat C^{1/2}\ , \ \ \ \ \ \ {\cal
B} = \hat C^{1/2} \hat B \hat C^{1/2}\ . \ee
Now, defining
\be {\cal F}(v) = (v^2 + {\cal B}^2)^{-1}f{\cal A}\ , \ \ \ \ \
{\cal G}(v) = (v^2 + {\cal B}^2)^{-1}f \hat  A \ , \ee
we can write
\be m_X^2 = M_1^2 - M_2^2\ , \label{EME}\ee
where
\begin{eqnarray}
M_1^2 &=& \frac{1}{16\pi^2 f^2} {\rm Tr} \int_0^\infty dv\ v^5
{\cal F}^2\frac{{\cal G}^2}{1 - {\cal G}^2}\label{Eme1Def}\\
M_2^2 &=& \frac{1}{2}\frac{1}{16\pi^2 f^2} {\rm Tr} \int_0^\infty
dv\ v^3 \left(\frac{{\cal G}}{1 - {\cal G}^2}\left\{{\cal F}, {\cal
B}\right\}\right)^2\ . \label{Eme2Def}
\end{eqnarray}

It is easy to see that if only fields with R-charge $R = 0, 2$
assignments are present, then $M_2^2 = 0$ and $M_1^2 > 0$, so
R-symmetry is not broken as it happens in \cite{Shih:2007av}. This
suggests that generic R-charge assignments should be made to quiral
models in order for the R-symmetry to be broken. In the next section
we consider a model of this kind.

Let us end this section with a brief disclaimer about explicit
R-symmetry breaking and some comments on the phenomenological consequence  of considering non-canonical K\"ahler in relation to R-symmetry.
Explicit R-symmetry in these models have been
studied in \cite{Intriligator:2007py}, \cite{Abe:2007ax}. In these
works the K\"{a}hler is canonical and R-symmetry breaking terms are
added to the superpotential, leading to the appearance of
supersymmetric vacua, in agreement with the Nelson-Seiberg argument
\cite{Nelson:1993nf}. As expected, in the limit of small R-symmetry
breaking the susy vacua can be pushed sufficiently far from the
origin of field space, thus making the metastable vacua
parametrically long-lived. Trying to repeat this procedure by
breaking R-symmetry from the K\"{a}hler potential fails. The reason for
this is that the conditions for supersymmetry breaking  depend only
on the superpotential $W$ and not on the form of the K\"{a}hler
potential.

As stated by Nelson and Seiberg \cite{Nelson:1993nf}, it is a necessary condition for SUSY-breaking in generic models to have an R-symmetry, and a sufficient condition that R-symmetry is spontaneously broken. From a phenomenological point of view this fact is problematic because an unbroken R-symmetry forbids Majorana gaugino masses, and having an exact but spontaneously broken R-symmetry leads to a light R-axion. Let us mention how we get rid of this apparent problem. First of all lets us comment that our model is not generic, so the Nelson-Seiberg argument is not applicable. Therefore, we could in principle be able to break R-symmetry explicitly from the superpotential without restoring SUSY (see an example in \cite{Abe:2007ax}). We have not explored this possibility, instead we have considered two other cases. One in which (following \cite{Shih:2007av}) we have an R-symmetry which can be spontaneously broken. In this case, we can expect that including gravity will make the R-axion sufficiently massive \cite{Bagger:1994hh}. Another case, commented in the previous paragraph, in which R-symmetry is explicitly broken in the K\"ahler potential. This possibility is free from the above problems, since R-symmetry breaking does not induce SUSY vacua, and we have no goldstone boson because the symmetry is explicitly broken. Another possibility for spontaneous R-symmetry breaking could be choosing an R-symmetric K\"ahler potential generating a tree-level lifting of the moduli space giving rise to a non-R-symmetric minimum.

\section{Shih model with non-canonical K\"{a}hler}

We have shown in the previous section that in order for the
O'Raifeartaigh models (\ref{KahlerC})-(\ref{SuperpotNM}) to have
R-symmetry broken, there must be in the theory at least one field
with R-charge different from $0$ or $2$. A model of this kind was
proposed in \cite{Shih:2007av}. The model has superpotential
\be W = \lambda X \phi_1 \phi_2 + m_1 \phi_1 \phi_3 + \frac{1}{2}
m_2 \phi_2^2 + f X\ , \ee
and in our case the K\"{a}hler potential reads
\be K = \bar X X + C(\bar \phi^{\bar j}, \phi_j) \ . \ee
By rotating the phases of all the fields, the couplings can be taken
to be real and positive, without loss of generality. In order for
the theory to be R-symmetric, the R-charge assignments must be $R(X)
= 2$, $R(\phi_1) = -1$, $R(\phi_2) = 1$ and $R(\phi_3) = 3$. Notice that because the R-charge assignments are all different, the K\"ahler
 potential depends on the fields in the form $\phi_i \bar \phi^i$. Then, the transformation that takes the quadratic part of the K\"ahler to its canonical form consists only of a rescaling of the fields. This rescaling, together with a redefinition of the constants, leaves the superpotential invariant. In other words, this model is (near $\phi = 0$) nothing but the Shih model with redefined constants. We review some properties of this model to see what can be changed by considering non-canonical K\"ahler.

The extrema of the potential consists of the pseudo-moduli space
\be \phi_i = 0 \ , \ \ \ \ \forall X\ \ \ \ \ \longrightarrow \ \ \
\  V_0 = f^2\ . \ee
This is the only extrema if the K\"{a}hler is canonical. Depending on
the explicit form of $C(\bar \phi^j \phi_j)$ other extrema can
appear. The pseudo-moduli space is a minimum of the potential when
\be |X| < \frac{c_1}{2}\ \left(1 -
\frac{f\lambda}{c_2c_3m_1m_2}\right)/\left(\frac{f\lambda^2}{m_1^2m_2}\right)\
, \ee
where
\be c_i \equiv \left(\frac{\partial C} {\partial (\phi_i \bar
\phi^i)}\right)^{-1}_{\phi = 0}\ , \ee
otherwise some eigenvalues of the mass squared matrix become
tachyonic. This pseudo-moduli space is only a local minima of the
potential provided there is a runaway direction
\be X = \left(\frac{m_1^2 m_2 \phi_3^2}{f \lambda^2 }\right)^{1/3}\
, \ \ \ \phi_1 = \left(\frac{m_2 f^2}{\lambda^2 m_1
\phi_3}\right)^{1/3}\ , \ \ \ \phi_2 = - \left(\frac{m_1 f
\phi_3}{\lambda m_2}\right)^{1/3}\ , \label{Direction}\ee
as long as (\ref{RunawayCond}) is satisfied
\be \lim_{\phi_3 \to \infty}\phi_3^{-2/3}\left(\frac{\partial^2
C}{\partial \phi_3
\partial \bar \phi^3}\right)^{-1}  = 0\ .\ee
Notice that the direction of the runaway in field space is the same as its canonical counterpart, but the value of the scalar potential evaluated on this direction is modified. This runaway direction can be parameterized by $X$, and along this
direction the potential takes the values
\be V_{RA} (|X|) =  \left(\frac{\partial^2 C}{\partial \phi_3
\partial \bar \phi^3}\right)^{-1}\!\!\!\!\!\!(|X|) \ \frac{m_1^2 m_2 f}{\lambda^2 |X|} \ . \ee
The value of $|X|$ for which $V_{RA} = V_0$ gives an estimate of the
vacuums life-time. It can be taken parametrically long-lived, and
moreover, the K\"{a}hler potential can change the lifetime.

The $A$, $B$ and $C^{-1}$ matrices in
(\ref{KMatriz}),(\ref{ABMatrices}) are
\be A = \left(\begin{matrix}0 & \lambda & 0
\\ \lambda & 0 & 0 \\ 0 & 0 & 0 \end{matrix}\right)\ ,  \ \ \ B = \left(\begin{matrix}0 & 0 &
m_1 \\ 0 & m_2 & 0 \\ m_1 & 0 & 0 \end{matrix}\right)\ ,  \ \ \
C^{-1} = \left(\begin{matrix} c_1 & 0 & 0
\\ 0 & c_2 & 0 \\ 0 & 0 & c_3 \end{matrix}\right)\ . \ee
As a test of the expressions (\ref{Eme1Def})-(\ref{Eme2Def}) we have explicitly evaluated them for this non-canonical model and further compared the results with the rescaled results of \cite{Shih:2007av}. The calculation with the formulas we derived is
\begin{eqnarray}
M_1^2  &=& \frac{c_1^2 c_2^2}{4\pi^2 f^2}\int^\infty_0 dv\ v^5 \
\frac{f^4\lambda^4}{(v^2 + \tilde m_1^2)(v^2 + \tilde
m_2^2)\left((v^2 + \tilde m_1^2)(v^2 + \tilde m_2^2) - f^2
\lambda^2 \right)}\\
M_2^2 &=& \frac{c_1^2 c_2^2}{2\pi^2 f^2}\int^\infty_0 dv\ v^3 \
\frac{f^4\lambda^4\tilde m^2_2}{\left((v^2 + \tilde m_1^2)(v^2 +
\tilde m_2^2) - f^2 \lambda^2 \right)^2}
\end{eqnarray}
where we defined $\tilde m_1 = c_1 c_3 m_1$ and $\tilde m_2 = c_2
m_2$, and this can be rewritten as
\begin{eqnarray}
M_1^2  &=& \frac{c_1^2 c_2^2\lambda^2 \tilde m_1^2 r^2 y^2
}{4\pi^2}\int^\infty_0 dv\ v^5 \ \frac{1}{(v^2 + 1)(v^2 +
r^2)\left((v^2 + 1)(v^2 + r^2) - y^2 r^2 \right)}\\
M_2^2  &=& \frac{c_1^2 c_2^2\lambda^2 \tilde m_1^2 r^4 y^2
}{2\pi^2}\int^\infty_0 dv\ v^3 \ \frac{1}{\left((v^2 + 1)(v^2 + r^2)
- y^2 r^2 \right)^2}
\end{eqnarray}
where we have defined $y = \frac{\lambda f}{\tilde m_1 \tilde m_2}$
and $r = \tilde m_2 / \tilde m_1$.
Integrating this expression to order ${\cal O}(y^2)$ we obtain
\begin{eqnarray}
M_1^2  &=& \frac{c_1^2 c_2^2\lambda^2 \tilde m_1^2 r^2 y^2
}{8\pi^2}\ \frac{r^4 - 4 r \log(r) - 1}{(r^2 - 1)^3} + {\cal O} (y^4) \\
M_2^2  &=& \frac{c_1^2 c_2^2\lambda^2 \tilde m_1^2 r^4 y^2
}{2\pi^2}\ \frac{(r^2 + 1)\log(r) + 1 - r^2}{(r^2 - 1)^3} + {\cal O}
(y^4)\ .
\end{eqnarray}
These expressions are identical to those obtained in
\cite{Shih:2007av}, although in this case the definition of the
parameters $y$ and $r$ depend on the K\"{a}hler potential. It was shown
in \cite{Shih:2007av} that some $r^*$ exists such that for $r > r^*$
the $m_X^2 = M_1^2 - M_2^2 < 0$, so R-symmetry is broken. A
non-canonical K\"{a}hler potential cannot change this behavior, although
it can change the value of $r^*$.

\section{Summary and Discussion}

Several aspects of R-symmetry and supersymmetry breaking have been
studied in generalized O'Raifeartaigh models with non-canonical
K\"{a}hler potential. We derived some conditions on the K\"{a}hler potential
in order for the non-supersymmetric vacua to be degenerate at
tree-level. This is a common feature of renormalizable models and we
show that it is also shared by many non-renormalizable theories.

Once degeneracy is guaranteed for the vacuum at the classical level,
the information about the lifting of the flat directions is given by
the CW effective potential. We calculated the CW potential for
arbitrary quiral non-linear sigma-models, and this allowed us to
study the 1-loop quantum corrections to the pseudo-moduli space.
This potential has a quadratic and a quartic dependence on the cutoff scale
$\Lambda$ which vanish identically in supersymmetric models with
canonical K\"{a}hler. In our case the quadratic dependence also vanishes, which can be seen as a consequence of R-symmetry in our model, and the quartic dependence becomes constant in the considered vacuum. Concerning the logarithmic divergent term
$\log(\Lambda)\ {\rm STr}\ {\cal M}^4$, it can usually be absorbed into
the renormalization of the coupling constants appearing in the
tree-level vacuum energy in theories with canonical K\"{a}hler. It would also be interesting to study if
this is the case in our non-renormalizable R-symmetric models.
Another interesting fact is that the mass of the flat mode is
independent of $\Lambda$ also due to R-symmetry, this happens in
renormalizable models as well. These similarities between
R-symmetric models with canonical K\"{a}hler potential, and R-symmetric
models with non-canonical K\"{a}hler potentials require further
research. One may wonder if these similarities between models with
canonical and non-canonical K\"{a}hler are extensive for any R-symmetric
superpotential, or if they are only valid in this generalized
O'Raifeartaigh model.

The conditions for R-symmetry breaking remain unchanged with respect to those of canonical models.
R-symmetry can be broken when generic R-charge assignments to the
fields are made, while R-symmetry remains unbroken when only fields
with R-charge $0$ and $2$ are present. In \cite{Ray:2007wq}, based
on the number of fields with $0$ and $2$ R-charge, more information
is obtained about the properties of the model regarding symmetry
breaking. It would be interesting to see if a similar analysis can
be done in the case of non-canonical models. Another issue to be
more thoroughly analyzed
concerns the question whether two K\"ahler potentials exist, such that for a fixed
superpotential, R-symmetry is broken in one case and unbroken in the
other.

The models we presented   can keep the runaway behavior of their canonical
counterparts. Moreover, non-minimal K\"{a}hler potentials can induce the
existence of new runaway directions. These directions imply that the
non-supersymmetric vacua is metastable, and the life-time of the
vacuum depends on the form of the K\"{a}hler potential.

\section*{Appendix: One-loop effective potential for non-linear sigma \-models.}

In this appendix we calculate the Coleman-Weinberg effective
potential \cite{Coleman:1973jx} for a sigma model with general
K\"{a}hler potential $K$ and superpotential $W$. In
\cite{Coleman:1973jx}, the computation is made for renormalizable
theories, so we must recalculate it. The model we consider has $N$
superfields $Z^a$, with scalar component $z^a$ and fermionic
component $\psi^a$. The action of the theory is
\be {\cal S} = \int d^4 x \left[\int d^2\theta d^2\bar\theta K(Z,
\bar Z) + \int d^2\theta W(Z) + \int d^2\bar\theta W(\bar Z)\right]\
.\ee

Recalling that for a K\"{a}hler manifold the covariant derivative, the
connection and the curvature take the form

In terms of the quantities
\begin{eqnarray}
    V&=& \bar W_{\bar a} K^{\bar a a} W_a\nonumber\\
    D_a W_b &=& \partial_a W_b - \Gamma^c_{ab} W_c\nonumber\\
    \Gamma^c_{ab} &=& K^{\bar c c}\ \partial_a\ K_{b \bar c}\nonumber\\
    D_\mu \psi^a &=& \partial_\mu \psi^a - \Gamma^a_{bc} \partial_\mu
    z_b \psi_c\nonumber\\
    (R_{\bar b b})^{a\bar a} &=& K^{\bar a c}\partial_{\bar b} \Gamma^{a}_{bc}\ ,
\end{eqnarray}
and after integrating the $\theta,\bar\theta$ variables and the
auxiliary fields, we obtain the action
\begin{eqnarray}
{\cal S} &=& \int d^4 x \left[K_{a\bar a} \left(\partial_\mu z^a
\partial^\mu \bar z^{\bar a} + \frac{i}{2} D_\mu \psi^a \sigma^\mu \bar \psi^{\bar a} - \frac{i}{2} \psi^a
\sigma^\mu D_\mu \bar \psi^{\bar a} \right) -
V(z^a, \bar z^{\bar a}) \right. \nonumber\\
&-& \left. \frac{1}{2} D_a W_b \psi^a \psi^b - \frac{1}{2} D_{\bar
a} \bar W_{\bar b} \bar \psi^{\bar a} \bar \psi^{\bar b} +
\frac{1}{4} R_{\bar a a \bar b b} \psi^a\psi^b \bar \psi^{\bar
a}\bar \psi^{\bar b}\right]
\end{eqnarray}
Being $z_0^a$ the VEV of the scalar fields, we define small
fluctuations $z^a \to z_0^a + \sqrt{\hbar} \varphi^a$ and $\psi_a
\to \sqrt{\hbar}\chi_a$, and to order ${\cal O}(\hbar)$ we are left
with
\begin{eqnarray}
{\cal S}^{(1)} &=& \int d^4 x \left[K_{a\bar a}(z_0)
\left(\partial_\mu \varphi^a
\partial^\mu \bar \varphi^{\bar a} + \frac{i}{2} \partial_\mu \chi^a \sigma^\mu \bar \chi^{\bar a}
- \frac{i}{2}\chi^a \sigma^\mu \partial_\mu \bar \chi^{\bar a}
\right) - \frac{1}{2} D_a W_b(z_0) \chi^a \chi^b - \frac{1}{2}
D_{\bar a} \bar W_{\bar b}(z_0) \bar \chi^{\bar a} \bar \chi^{\bar b} \right. \nonumber\\
&-& \left. \frac{1}{2}\left(\partial_{a}\partial_{ b} V (z_0)
\varphi^{a} \varphi^{b} + \partial_{\bar a}\partial_{\bar b} V (z_0)
\bar\varphi^{\bar a} \bar \varphi^{\bar b} + 2
\partial_a\partial_{\bar b} V (z_0) \varphi^a \bar \varphi^{\bar b}\right)\right]\
,
\end{eqnarray}

In terms of the $N$ scalar fields $\Phi^a$ and the $N$ Dirac spinors
$\Psi^a$ given by
\begin{eqnarray}
\Phi^a  = \left(\begin{matrix}\varphi^a\\
\bar\varphi^a\end{matrix}\right)\ , \ \ \ \ \ (\Phi^a)^\dag =
\left(\bar\varphi^a\ \ \ \varphi^a\right)\ , \ \ \ \ \ \Psi^a  = \left(\begin{matrix}(\chi^a)_\alpha\\
(\bar\chi^a)^{\dot \alpha}\end{matrix}\right)\ , \ \ \ \ \ \bar
\Psi^a = -(\Psi^a)^\dag \gamma^0 = ((\chi^a)^\alpha \ \ \ (\bar
\chi^ a)_{\dot\alpha})\ ,
\end{eqnarray}
where we choose the Weyl basis for the $\gamma$-matrices
\be \gamma^\mu = \left(\begin{matrix} 0_2 & \sigma^\mu \\
\bar \sigma^\mu & 0_2\end{matrix}\right)\ , \ \ \ \ \sigma^\mu =
(-1, \sigma^i)\ , \ \ \ \ \bar\sigma^\mu = (-1, - \sigma^i)\ ,\ee
we can rewrite ${\cal S}^{(1)}$ as
\begin{eqnarray}
{\cal S}^{(1)} &=& \frac{1}{2} \int d^4 x \left[
\partial_\mu (\Phi^a)^\dag {\cal K}^B_{ab}\partial^\mu \Phi^b - (\Phi^a)^\dag ({\cal M}_B^2)_{ab} \Phi^b  -i
\bar \Psi^a {\cal K}^F_{ab} \gamma^\mu \partial_\mu \Psi^b
-\bar\Psi^a (M_F)_{ab} \Psi^b\right]
\end{eqnarray}
where the matrices ${\cal K}^B_{ab}$ and ${\cal K}^F_{ab}$ are defined by
\bea {\cal K}_{ab}^B = \left(\begin{matrix} K_{ba}&  0\\
                                          0 & K_{ab}
\end{matrix}\right)\,\quad\quad\quad{\cal K}^F_{ab} = \left(\begin{matrix} K_{ab}  {\bf 1}_2&  0\\
                                          0 & K_{ba}{\bf 1}_2
\end{matrix}\right)= {\cal K}_{ba}^B\otimes {\bf 1}_2
 \
\eea
while the mass matrices for bosons and fermions can be written as

\be {\cal M}^2_B = \left(\begin{matrix} D_{\bar b}\bar W_{\bar a}
K^{\bar a a} D_b W_a - \bar W_{\bar a} (R_{\bar b b})^{a\bar a} W_a
& \partial_{\bar b\bar c}(\bar W_{\bar a} K^{\bar a a}) W_a
\\ \bar W_{\bar a} \partial_{bc}(K^{\bar a a} W_a) &
D_{\bar b}\bar W_{\bar a} K^{\bar a a} D_b W_a - \bar W_{\bar a}
(R_{\bar b b})^{a\bar a} W_a \end{matrix}\right)\ , \label{MB2}\ee
\be {M}_F = \left(\begin{matrix}D_b W_a{\bf 1}_2& 0 \\ 0 & D_{\bar
b}\bar W_{\bar a}{\bf 1}_2
\end{matrix}\right)\ .\label{MF2}
\ee
Therefore, the first order correction to the effective potential
reads
\be V^{(1)}_{\rm eff} = - \log\left({\det}^{-\frac 12}\left(\hat {\cal
B}\right)\ {\det}^{\frac 12}\left(\hat {\cal F}\right) \right)\ee
where we have introduced the operators
\be \hat{\cal B}_{ab} = - {\cal K}^B_{ab}\square - ({\cal
M}_B^2)_{ab} \ , \ \ \ \ \ \ \hat{\cal F}_{ab} = \gamma^0 (-i{\cal
K}^F_{ab}\ \not\!\partial - (M_F)_{ab})\ . \ee
After passing to momentum space, the one-loop correction to the
potential reads
\be V^{(1)}_{\rm eff} = \frac 1 2 Tr \int \frac{d^4 p}{(2\pi)^4}
\left[\log({\cal K}^B_{ab}p^2 - ({\cal M}_B)_{ab}^2) -
\log(\gamma^0(- {\cal K}^F_{ab} \not\!p -(M_F)_{ab})) \right] . \ee
Introducing the $2\times 2$ mass matrix for fermions ${\cal M}_F$ through the definition
\be {\cal M}_F \otimes {\bf 1}_2 = - \gamma^0 M_F\ ,\ee
and using the properties
\be \gamma^0 {\cal K}^F_{ab}\gamma^0 = {\cal K}^{F \ \dag}_{ab} =
{\cal K}^B_{ab} \otimes {\bf 1}_2 \ ,\quad\quad\quad [{\cal K}^B_{ab} \otimes {\bf 1}_2, \gamma^0 \gamma^\mu] =
0 \ee
we can rewrite the expression for $V^{(1)}_{\rm eff}$ as
\be V^{(1)}_{\rm eff} = \frac 1 2 Tr \int \frac{d^4 p}{(2\pi)^4}
\left[-\log({\cal K}^B_{ab}) + \log(p^2{\bf 1}_2 + \tilde{{\cal
M}_B^2}) -  \log( - \gamma^0 \not\!p - \tilde{\cal M}_F \otimes
1_2) \right] \ee
where
\begin{eqnarray}
\tilde {\cal M}_B^2 &=& {\cal K}_B^{-1/2} {\cal M}_B^2
{\cal K}_B^{-1/2}\nonumber\\
\tilde {\cal M}_F &=& {\cal K}_B^{-1/2} {\cal M}_F {\cal
K}_B^{-1/2}\ .
\end{eqnarray}
As usual one can express the trace of the Dirac operator as the trace of a Klein-Gordon operator, i.e.
\be
Tr\! \!\int\! {d^4 p}\,\log( - \gamma^0 \not\!p - \tilde{\cal M}_F \otimes
1_2) = 2 \,Tr\!\! \int\! {d^4 p}\,\log( p^2 {\bf 1}_2 - \tilde{\cal M}_F^2 )
\ee
which yields the following form for the one-loop correction to the
potential in Euclidean signature
\be V^{(1)}_{\rm eff} = \frac12 Tr \int \frac{d^4 p}{(2\pi)^4}
\left[-\log({\cal K}^B) + \log(p^2 {\bf 1}_2 + \tilde{{\cal M}_B^2}) -
\log( p^2 {\bf 1}_2 + \tilde{\cal M}_F^2 ) \right] \ee
Finally, using that $d^4 p = p^3\ dp\ d\Omega$, $\int d\Omega =
2\pi^2$, and
\be \int dp p^3 \log[p^2 + m^2] = \frac{1}{4}(p^4 - m^4)\log(p^2 +
m^2) + \frac{p^2 m^2}{4} - \frac{p^4}{8} \ ,\ee
we obtain the desired formulae after cutoff regularization
\bea V^{(1)}_{\rm eff} &=& \frac{1}{64 \pi^2} {\rm Tr}\left(
\tilde{\cal M}_B^4 \left[\log\left(\frac{\tilde {\cal
M}_B^2}{\Lambda^2}\right) - \frac{1}{2}\right] - \tilde {\cal M}_F^4
\left[\log\left(\frac{\tilde {\cal M}_F^2}{\Lambda^2} \right) -
\frac{1}{2} \right]\right.\nonumber\\
&&\quad \left.
 + \frac{\Lambda^2}{2} \left(\tilde {\cal M}_B^2
- \tilde {\cal M}_F^2\right) - \Lambda^4 \log({\cal K}^B)\vphantom{\frac{\tilde {\cal
M}_B^2}{\Lambda^2}}\right) \ . \label{EffectivePot}\eea

\vspace{1 cm} \noindent\underline{Acknowledgements}: The authors wish to
thank Fidel A. Schaposnik for constant encouragement and useful discussions
during this work. This work was partially supported by UNLP, UBA, CICBA,
CONICET and ANPCYT.
\newpage

\end{document}